\documentclass[lettersize,journal]{IEEEtran}
\usepackage{amsmath,amsfonts}
\usepackage{algorithmic}
\usepackage{algorithm}
\usepackage{array}
\usepackage[caption=false,font=normalsize,labelfont=sf,textfont=sf]{subfig}
\usepackage{textcomp}
\usepackage{stfloats}
\usepackage{url}
\usepackage{verbatim}
\usepackage{graphicx}
\usepackage{cite}
\usepackage{tikz}
\begin{document}

\title{Tuning the Weights: The Impact of Initial Matrix Configurations on Successor Features  Learning Efficacy}

\author{Hyunsu Lee~\IEEEmembership{}
\thanks{H. Lee is with the Faculty of Department of Physiology, School of Medicine, Pusan National University, Yangsan, Republic of Korea.}
\thanks{}}

\markboth{H. Lee: weight analysis of SF agents}%
{H. Lee: weight analysis of SF agents}


\maketitle

\begin{abstract}
The focus of this study is to investigate the impact of different initialization strategies for the weight matrix of Successor Features (SF) on learning efficiency and convergence in Reinforcement Learning (RL) agents. Using a grid-world paradigm, we compare the performance of RL agents, whose SF weight matrix is initialized with either an identity matrix, zero matrix, or a randomly generated matrix (using Xavier, He, or uniform distribution method). Our analysis revolves around evaluating metrics such as value error, step length, PCA of Successor Representation (SR) place field, and the distance of SR matrices between different agents. The results demonstrate that RL agents initialized with random matrices reach the optimal SR place field faster and showcase a quicker reduction in value error, pointing to more efficient learning. Furthermore, these random agents also exhibit a faster decrease in step length across larger grid-world environments. The study provides insights into the neurobiological interpretations of these results, their implications for understanding intelligence, and potential future research directions. These findings could have profound implications for the field of artificial intelligence, particularly in the design of learning algorithms.
\end{abstract}


\section{Introduction}
\IEEEPARstart{F}{or} survival, animals are compelled to explore and interact with their environments. This interaction is underpinned by the ability to remember details about the environment, which enables animals to form expectations about future events or states based on their decisions. This capacity to predict outcomes based on past experiences is a cornerstone of intelligence. In animal cognition, the hippocampal system governs these predictive capabilities and memory functions \cite{andersen2006:hpc}.

The activity of \textit{place cells} in the hippocampus has long been of interest in the context of learning and memory. These specialized neurons, integral to the brain's limbic system, activate when an animal finds itself in a particular location \cite{okeefe1971, okeefe1976}. They essentially form a cognitive map or a neural embodiment of the spatial environment, which is pivotal for memory formation and learning. The discovery of place cells has led to numerous theories attempting to elucidate their role and the overarching function of the hippocampus in comprehending and learning spatial information. 

Among a myriad of theoretical models, the successor representation (SR) has proven to be an influential explanation for the role of the hippocampus in spatial representation \cite{stachenfeld2017}. SR posits that the hippocampus forms a cognitive map that is not a static spatial representation but rather a dynamic, anticipatory map that predicts future locations based on the current state \cite{geerts2020, stachenfeld2017}. The predictive map theory, interpreting place cell activity through the lens of SR learning, has shown considerable explanatory power for in vivo place cell activity. When an animal is first exploring an environment, its movements are random, and the expectation pattern appears symmetrical in all directions. As an animal becomes familiar with its environment, the activity of its place cells changes. While place cell activity exhibits a geodesic pattern during initial exploration, it transitions to an asymmetrical firing pattern as the animal becomes accustomed to the environment \cite{mehta2000}. 

According to the predictive map theory, the change in the firing pattern can be attributed to a shift in response from the act of visiting a specific location to the expectation of visiting that location. This shift in place cell activity results in a pattern that leans toward the animal's direction of movement because the expectation increases as the animal nears the location. The SR theory goes beyond predicting the immediate subsequent state, suggesting that the hippocampus forecasts all future states. This ability to construct a predictive map of the environment encompasses the animal's anticipations of future states, given its current state and behavior. This model elegantly bridges the gap between spatial navigation and reinforcement learning \cite{decothi2020, george2023, fang2023, bono2023}.

While SR provides a compelling explanation for place cell activity patterns, it presumes the animal has comprehensive knowledge of the environment's size and fully observable location information. To overcome this limitation and extend the predictive theory to partially observable environments, recent studies have proposed a feature-based SR model as a representation of hippocampal activity \cite{geerts2020, barreto2017}.

\IEEEpubidadjcol

In contrast to the traditional SR learning, which employs tabular methods and views each state as a separate entity, feature-based SR---also known as successor feature (SF)---employs a neural network as a function approximator to learn SR \cite{barreto2017}. This adaptation equips SR learning with the capacity to manage high-dimensional state spaces, making SF a more plausible neurobiological model than its naive counterpart. Nevertheless, the approach to initialize the weight matrix of the neural network varies significantly across the literature \cite{barreto2017,geerts2020}. Synaptic weights play a crucial role in neural networks, affecting the speed and success of learning. Despite its crucial role, the influence of different weight matrix initialization methods on overall learning remains largely unexplored.

In this research, our objective is to investigate to discern the influence of varying synaptic weight initialization patterns on SF learning. By subjecting SF learners to a basic maze environment under differing weight initialization patterns, we aim to illuminate the role of weight initialization in the SF learning process. For the evaluation of these impacts, we conducted an experiment utilizing identity, zero, and random matrices for weight initialization. With an $\epsilon$-greedy policy, the performances of the SR agent and the non-random SF agents were observed to be comparable in a one-dimensional (1D) maze. However, SF agents with randomly initialized weight matrices exhibited superior performance compared to their non-random counterparts. In the results section, we delve into the changes in the SR matrix throughout the learning process. Furthermore, in the discussion section, we reflect upon the neurobiological implication of weight matrix initialization. This investigation contributes to the continuous pursuit of understanding intelligence from both neuroscientific and artificial intelligence viewpoints.

\section{Model and Methods} \label{sec:methods}

\subsection{Successor Representation (SR)} \label{sec:SR}

In this study, we assume that an RL agent interacts with the environment through \textit{Markov decision processes} (MDP, \cite{puterman2014, sutton2018a}). An MDP is a tuple $M := (\mathcal{S}, \mathcal{A}, R, \gamma )$ comprising of the following elements. Sets $\mathcal{S}$ and $\mathcal{A}$ are the state (e.g., spatial locations) and action spaces. The function $R(s)$ specifies the immediate reward received in state $s$, which can be expressed as $R: \mathcal{S} \rightarrow \mathbb{R}$. Here, the discount factor $\gamma \in [0,1) $ is a weight that reduces the reward in the distant future. 

In RL, the agent's objective is to discover a policy function $\pi: \mathcal{S} \mapsto \mathcal{A}$ that maximizes the cumulative discounted reward, often referred to as the return $G_{t} = \sum^{\infty}_{i=t}{\gamma^{i-t}R_{i+1}}$, where $R_{t} = R(S_{t})$. The return essentially represents the sum of all future rewards that an agent can expect to accumulate, discounted by the factor $\gamma$. In order to solve this optimization problem, a common approach is to employ \textit{dynamic programming}, which defines and computes the \textit{value function} of a policy $\pi$ as follows:
	\begin{equation} \label{value_vanila}
	    V^{\pi}(s) := \mathbb{E}^{\pi}[G_{t} | S_{t} = s ]
	\end{equation}
where $\mathbb{E}^{\pi}[\cdot]$ denotes the expected value when the agent follows policy $\pi$. After determining $V^{\pi}(s)$, also known as \textit{policy evaluation}, the policy $\pi$ can be improved in a greedy manner. This process, referred to as \textit{greedy policy improvement}, is defined as follows: $\pi'(s) \in \text{argmax}_{a}Q^{\pi}(s, a)$ where $Q^{\pi}(s,a) := \mathbb{E}[R_{t+1} + \gamma V^{\pi}(S_{t+1}) | S_{t} = s, A_{t} = a]$ \cite{sutton2018a}. Here, $Q^{\pi}(s,a)$ represents the expected return from taking action $a$ in state $s$ and following policy $\pi$ thereafter.

As proposed in literature \cite{dayan1993}, the central premise of SR learning lies in the decomposition of the value function (Eq. \ref{value_vanila}). It suggests that the value function can be decomposed into an expected visiting occupancy and reward of the successor states $s'$ as follows:	
	\begin{equation} \label{value_decomposed}
	\begin{split}
		V^{\pi}(s) &= \sum_{s'}\mathbb{E}^{\pi}[\sum_{i=t}^{\infty}\gamma^{i-t}\mathbb{I}(S_{i} = s')R(s') | S_{t} = s] \\
		 &= \sum_{s'}M(s, s')R(s')
	\end{split}	
	\end{equation}
where $\mathbb{I}(S_{i}=s')$ yields a value of 1 when an agent visits the successor state $s'$ at time $t$; otherwise, it returns 0. Consequently, $M(s, s')$ represents the discounted expectation of visitation to the successor state $s'$ from the state $s$. $M(s, s')$ can be perceived as a comprehensive representation that integrates not only the immediate transition probabilities from state $s$ to state $s'$, but also the cumulative impact of the agent's policies and the array of potential future trajectories. This interpretation underscores the dynamism and predictive capacity of the SR, as it encapsulates the influence of the agent's decisions and environmental dynamics on future state visitations \cite{dayan1993}.

The SR matrix $M$ can be incrementally learned by the agent through the use of the temporal difference (TD) learning algorithm. This approach allows the agent to continually update its understanding of the environment based on the difference between predicted and actual visitation. The specific TD learning equation for the SR matrix $M$ is derived as follows \cite{stachenfeld2017,dayan1993}:
  
	\begin{equation} 
		\label{update_M}
		\Delta M(s_{t}, s') = \alpha_{M}[ \mathbb{I}(s_{t} = s') +  \gamma M(s_{t+1}, s') - M(s_{t}, s') ]  \
	\end{equation} 
	
\subsection{Feature-based SR}

The classical form of SR learning is constrained to tabular environments, limiting its applicability to more complex, high-dimensional settings \cite{lee2020a}. An effective means of circumventing this limitation is the application of a set of feature functions, denoted as ${\psi(s)}$, which allows for the generalization of SR learning \cite{barreto2017}. 

By assuming the expected reward of state $s$ can be represented as a product of the feature vector and its corresponding reward weights, denoted as $R(s) = \phi(s)^{T}\bold{w}_{rew}$, we can reframe the value function (Eq. \ref{value_vanila}) in a way that accommodates these feature functions. The revised value function is given as follows:
	\begin{equation}\label{successor_feature}
		\begin{split}
		V^{\pi}(s) &= \mathbb{E}^{\pi}[\sum_{i=t}^{\infty}\gamma^{i-t}\phi_{i+1}|S_{t} = s]^{T}\bold{w}_{rew}	\\
		& := \psi^{\pi}(s)^{T}\bold{w}_{rew}	
		\end{split}
	\end{equation}
where $\phi_{t}$ denote $\phi(S_{t})$. By incorporating a one-hot vector in $\mathbb{R}^{|\mathcal{S}|}$ for a tabular environment, $\psi(s)$ essentially mirrors the $M(s,:)$ vector of SR learning.   This is because it represents the discounted sum of occurrences of $\phi(s')$ when a transition unfolds under policy $\pi$. For clarity, we will henceforth refer to $\psi^{\pi}(s)$ as the successor features (SF) associated with state $s$ under policy ${\pi}$.

The introduction of SF marks a significant broadening of the SR learning framework, facilitating its application across a wider spectrum of MDP environments, such as partially observable MDPs and those characterized by continuous states \cite{vertes2019}.

In our approach, the SFs are approximated utilizing a linear function represented as follows:

	\begin{equation} \label{SF_approximator}
		\hat{\psi}(s) = W^{T}\phi(s)
	\end{equation}

This estimation leans on the presumption that $\phi(s)$ operates as a population vector of neurons that responds to the state $s$ observed by an agent. The utilization of a linear function aligns with neurobiological models of hippocampal place cells and finds support in the literature, reinforcing its relevance and applicability in our research \cite{geerts2020, lee2020a,stachenfeld2017}.

To estimate $\psi(s)$, we apply the TD learning to update the weight matrix $W$. This procedure parallels the matrix $M$ updating method observed in successor representation (SR) learning, thereby offering a streamlined approach to SF estimation in reinforcement learning contexts.

	\begin{equation} \label{update_W}
		\Delta W = \alpha_{W}[\phi(s_{t}) + \gamma\hat{\psi}(s_{t+1}) - \hat{\psi}(s_{t})]\phi(s_{t})^{T}
	\end{equation}
	
It's worth highlighting that Eq. (\ref{update_W}) corresponds to Eq. (\ref{update_M}) when $\phi(s)$ is presented as a one-hot vector. Alongside this, the expectation weight vector associated with rewards, denoted as $\bold{w}_{rew}$, is updated using a simple delta rule as follows:

	\begin{equation} \label{rew_update}
		\Delta \bold{w}_{rew} = \alpha_{r}( R_t - \phi(s)^{T}\bold{w}_{rew})\phi(s)
	\end{equation}

With the established update rules, we are now ready to investigate the SFs' learning with different initialization methods of the weight matrix $W$. Notably, the initial values of weight matrix $W$ are assumed to play a critical role in the learning performance and efficiency.

\subsection{SF Leaners and Their Weight Initializatoin Patterns}
To explore the influence of different weight initialization methods on the learning dynamics of SFs, we initialized the weight matrix $W$ using three different methods: identity, zero, and small random matrices. 

\subsubsection{Identity matrix initialization}

The identity matrix initialization method sets the initial weight matrix $W$ as an identity matrix, $W = I$. This means that the initial estimates of the SFs are equivalent to the one-hot encoded state representations. This initialization strategy can be regarded as a "knowledgeable initialization," endowing the agents with preliminary information about the environment \cite{geerts2020}.

\subsubsection{Zero matrix initialization}
In contrast, the zero matrix initialization method sets all elements of the initial weight matrix $W$ to zero. This means the SFs initially predict no future state visitations, assuming no knowledge of the world at the initial state. This initialization method can be seen as a "naive initialization", where agents start learning from scratch without any prior knowledge about the environment.

\subsubsection{Small random matrix initialization}
Small random matrix initialization, a commonly employed method in machine learning, sets the initial weight matrix $W$ with small random values drawn from specific distributions. This technique infuses randomness into the preliminary estimates of SFs, conjecturing a mixture of accurate and imprecise understanding of the world at the onset \cite{barreto2017}. We employ a single layer for the successor feature. Given that the expected future visitation can't be negative, we ensure that the weights are initialized randomly within the positive domain by applying an absolute value function. For this investigation, we utilized three prevalent techniques to initialize small random matrices: the Xavier method, the He method, and a uniform distribution.

\paragraph{Xavier method}
The Xavier method, also known as Glorot initialization \cite{glorot2010}, is a popular method for weight initialization in deep neural networks. This method determines initial weights by drawing a random number from a uniform probability distribution ($U$) within the range of $\frac{-1}{\sqrt{n}}$ to $\frac{1}{\sqrt{n}}$. In our study, 'n' corresponds to the number of input neurons, thereby representing the size of the one-dimensional (1D) grid world. 

\paragraph{He method}
The He initialization technique \cite{he2015}, another approach utilized in this research, derives initial weights from a Gaussian probability distribution characterized by a mean of zero and a standard deviation given by $\sqrt{2/n}$, where 'n' symbolizes the number of input neurons.

\paragraph{Uniform distribution}
The Uniform distribution method represents the most straightforward approach to initializing small random matrices. In our study, this method involved distributing weights uniformly across an interval ranging from 0 to 0.1. This choice of distribution infers that we hold the expectation of future visitation to each successor state as uniformly probable.

\subsection{Experimental Set-up}

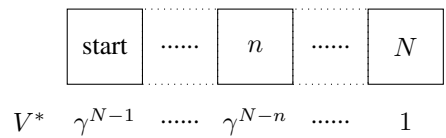
\begin{figure}[htpb]
		\centering
    	\begin{tikzpicture}

    		\draw node[rectangle, draw, minimum size=1cm] at (0,0) {start};
        	\draw node[rectangle, dotted, draw, minimum size=1cm] at (1,0) {......};
        	\draw node[rectangle, draw, minimum size=1cm] at (2,0) {$n$};
        	\draw node[rectangle, dotted, draw, minimum size=1cm] at (3,0) {......};
        	\draw node[rectangle, draw, minimum size=1cm] at (4,0) {$N$};
        	\node at (-1, -1){$V^{*}$};
        	\node at (0, -1){$\gamma^{N-1}$};
        	\node at (1, -1){......};
        	\node at (2, -1){$\gamma^{N-n}$};
        	\node at (3, -1){......};
        	\node at (4, -1){$1$};
    		
	    \end{tikzpicture} 
       
    	\caption{Schematic of the 1D grid world following the MDP. $V^{*}$ represents the expected true value of each cell according to discount factor ($\gamma$) when the reward of the terminal state is one.}
	    \label{fig1:GridWorld}
    \end{figure}

To investigate the learning process of each RL agent, we used a simple one-dimensional (1D) grid world of size $N \in \mathbb{N}$ spanning from 3 to 100 cells. In this environment, the agent navigates the grid world using left and right actions (Figure \ref{fig1:GridWorld}). In every episode, the agent starts at the leftmost position in the grid world. The ultimate goal is to reach the rightmost end of the world (also known as the terminal state). Upon reaching this terminal state, the agent receives a reward of 1 point. In contrast, all other states receive a score of 0, which means no reward. In our investigation, the discount factor $\gamma$ was set to 0.95.

To maximize the overall discounted reward, the agent selects actions predicated upon the estimated Q value utilizing an $\epsilon$-greedy policy. This policy prescribes a uniform random action selection with probability $\epsilon$, while at other times, with a probability of $1-\epsilon$, the agent chooses the action associated with the highest Q-value estimate. To foster adequate exploration and promote learning stabilization, the probability $\epsilon$ undergoes a decay according to the rule: $\epsilon_{k} = 0.9 \cdot 0.95^{k} + 0.1$, where $k$ signifies the episode index \cite{lehnert2017}.

The learning rates assigned to each learner---the matrix $M$ for the SR agent and the matrix $W$ for the SF agents—--were uniformly set at $\alpha_{M} = \alpha_{W} = 0.1$. The learning rate allocated to the reward position vector, for both the SR agent and the SF agents, was fixed at $\alpha_{r} = 0.1$. To equitably compare SR and SF agents, maze environment observations were utilized as state indices for the SR agent and one-hot coding vectors for the SF agents.

\subsection{Performance Evaluation Metrics}

We evaluated the learning performance of SF agents with different weight initialization methods based on several metrics, including learning speed, final performance, and stability of learning. In addition to these performance metrics, we also analyzed the changes in the SR matrix and the weight matrix throughout the learning process. These analyses allowed us to better understand the dynamics and mechanisms underlying the influence of weight initialization on SF learning. 

In this evaluation, each agent simulation test was run 10 times, and the mean and standard deviation of the results are presented in the experimental result section.

\subsubsection{Evaluating the evolution of SR place field matrix}

In an effort to elucidate the intricacies of the SR matrix's evolution and convergence patterns over the progression of episodes, we utilized Principal Component Analysis (PCA)---a powerful dimensionality reduction tool \cite{jolliffe2016}. This process was complemented by calculating the L1 distance between matrices at various stages throughout the learning episodes. This measurement helped in detailing the patterns of convergence inherent to the SR matrix as the agent gained expertise within the simple maze environment. This measurement was computed as follows:

\begin{equation}
\label{eq:L1dist}
d_{1}(\boldsymbol{A},\boldsymbol{B}) = \sum_{i=1}^{N} \sum_{j=1}^{N} |a_{ij} - b_{ij}|
\end{equation}

In this formula, $a_{i,j}$ and $b_{i,j}$ denote individual elements within the SR place field matrices $\boldsymbol{A}$ and $\boldsymbol{B}$, respectively. This methodological approach offers a comprehensive portrayal of the SR matrix's conversion throughout the unfolding learning episodes.

\subsubsection{Value error} \label{sec:eval}

Within a one-dimensional maze that begins from the leftmost position, the optimal policy would invariably guide movements towards the right. Accordingly, the legitimate value of the n-th grid cell, denoted as $V^{}(s_{n})$, amounts to $\gamma^{N-n}$, with $s_{n}$ representing the n-th grid cell. This investigation entailed a comparison of the learning efficiency among diverse agents, relying on the mean square error (MSE) as a metric. The MSE is the discrepancy between the true value function, $V^{*}$, and the value function under the current policy, $V^{\pi}$. It is mathematically represented as follows:

	\begin{equation} \label{V_error}
		\textrm{MSE} = \sum_{n=1}^{N-1}(V^{*}(s_{n})-V^{\pi}(s_{n}))^{2}
	\end{equation}
	
Apart from the aforementioned metric, we also incorporated an alternative measure, defined as $-\frac{\Delta\textrm{MSE}}{\Delta\textrm{episode}}$. This metric specifically provides insight into the rate at which the value error diminishes over time, hence offering an additional perspective on learning efficiency.

\subsubsection{Step length}

In our analysis, we utilized additional metrics to understand the agent's learning progression in a comprehensive manner. The step length, representing the number of steps the agent takes to reach the goal, was one such measure. As the agent improves its policy with learning, this step length is expected to decrease. We assessed the rate of change of the step length over episodes, defined as $\frac{\Delta(\textrm{step length})}{\Delta\textrm{episode}}$, to gain insights into the speed of the agent's learning and the pace of policy improvement. 

Further, we also evaluated the variability in the rate of step length decrease to understand the stability of learning. This was accomplished by calculating the standard deviation of $\frac{\Delta(\textrm{step length})}{\Delta\textrm{episode}}$ over the course of learning episodes. This metric allows us to gauge the consistency of the agent's learning progress, providing a more complete picture of the learning dynamics.

\section{Experimental results}

	\begin{figure*}[!ht]
		\centering
		\includegraphics[width=7in]{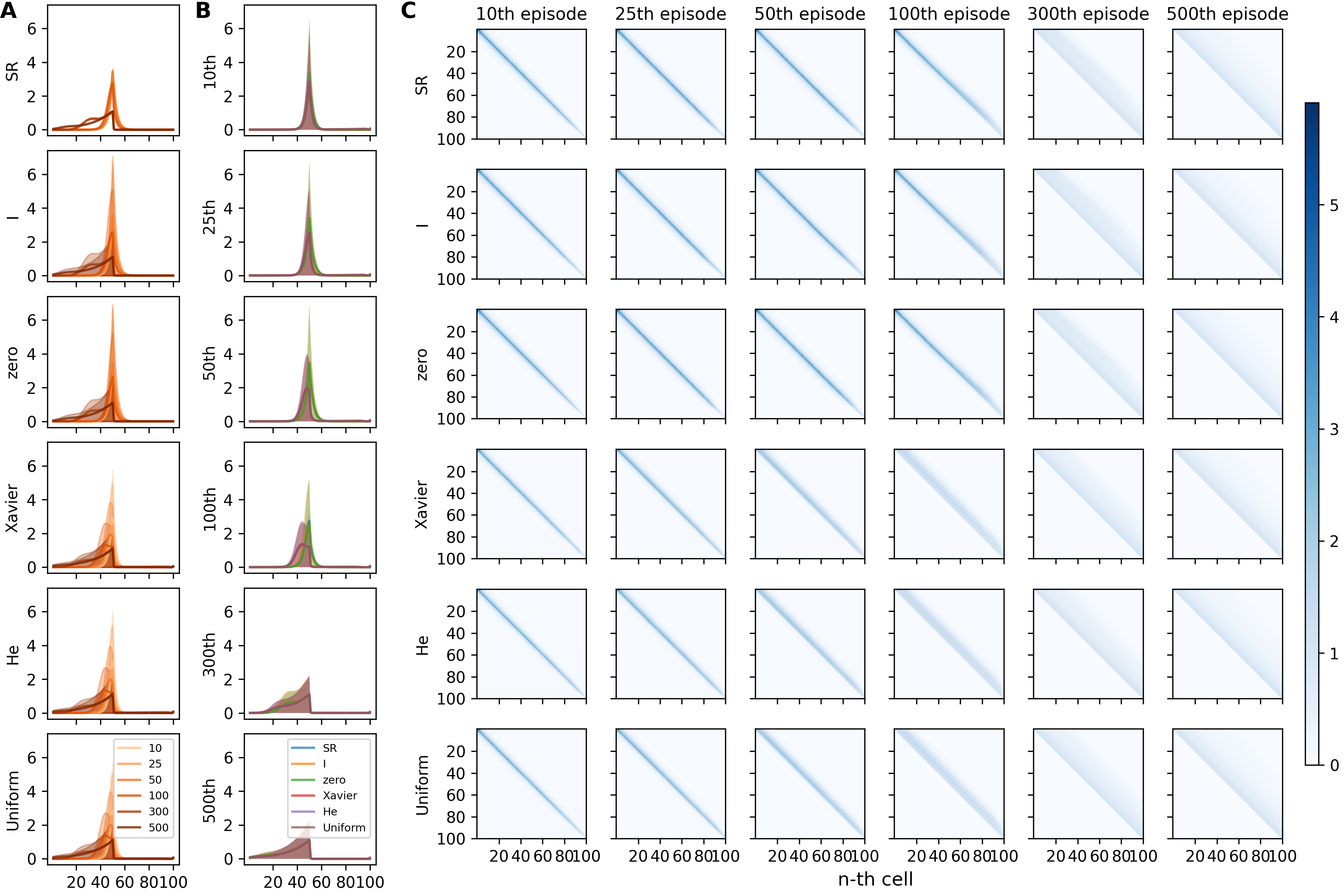}
		
		\caption{The simulated learning histories of the SR place field show that the SF agents, with the initial weight set with random weights, rapidly converges to the asymmetric SR place field. (A) Line plots of the learned SR place field of 50th cell after end of 10th, 50th, 100th, 300th episode are shown. Each line and shade show the averaged result with the standard deviation from 10 simulations in a grid world with 100 cells. Each row panel displays the SR agent (first row) or SF agents with different weight initialization methods (five rows below). (B) Rearranged line plots from A for comparing between the agents (SR, blue; SF weight initialization with identity matrix, orange; zero matrix, green; the Xavier method, red; the He method, purple; the uniform distribution, brown). Each row panel displays the simulated results after end of 10th, 25th, 50th, 100th, 300th, 500th episode. Note that the SF agents with random weights shows skewed SR place field at 50th episode, but other agents show symmetrical SR place field. (C) Learning histories of whole SR place field matrix in a grid world with 100 cells are shown according to episodes (column panels) and learning agents (row panels).}
	    \label{SR_matrix}
	\end{figure*}
	
\subsection{Accelerated Convergence of Random SF Agents to Asymmetrical SR Place Fields}

To elucidate the impact of disparate initial matrix forms on SF agents, we analyzed the transformational learning history of the weight variables of SF agents, juxtaposing the findings with those of the SR agent.

\subsubsection{Learning history of SR place field}
In Figure \ref{SR_matrix}, we present characteristic results simulated in a grid world comprising 100 cells. Upon comparison of the learning pattern of the 50th cell's SR place field across agents, we noticed that SF agents equipped with random weights (Xavier, He, uniform) exhibited an expedited shift towards asymmetrical SR place fields relative to their non-random counterparts (SR, Identity matrix, Zero matrix). 

In specific, the SR place fields of the 50th cell for non-random agents retained a symmetrical pattern even at the 50th and 100th episodes (Figure \ref{SR_matrix}B). In contrast, when we inspected the learned pattern of the comprehensive SR matrix at the 50th episode (Figure \ref{SR_matrix}C), it became evident that the SR place fields of random agents already displayed an asymmetrical pattern. This held true even for cells located proximally to the first cell. On the other hand, non-random agents, with the exception of those in the vicinity of the goal location, continued to exhibit a symmetrical pattern in their SR place fields. 

These observations underscore the intriguing finding that SF agents with random initial weights converge more rapidly towards asymmetrical SR place fields compared to non-random agents. This facet is especially pronounced in the early stages of learning, which can have implications on the temporal dynamics of learning and overall task performance.

	\begin{figure}[ht]
		\centering
		\includegraphics[width=3.5in]{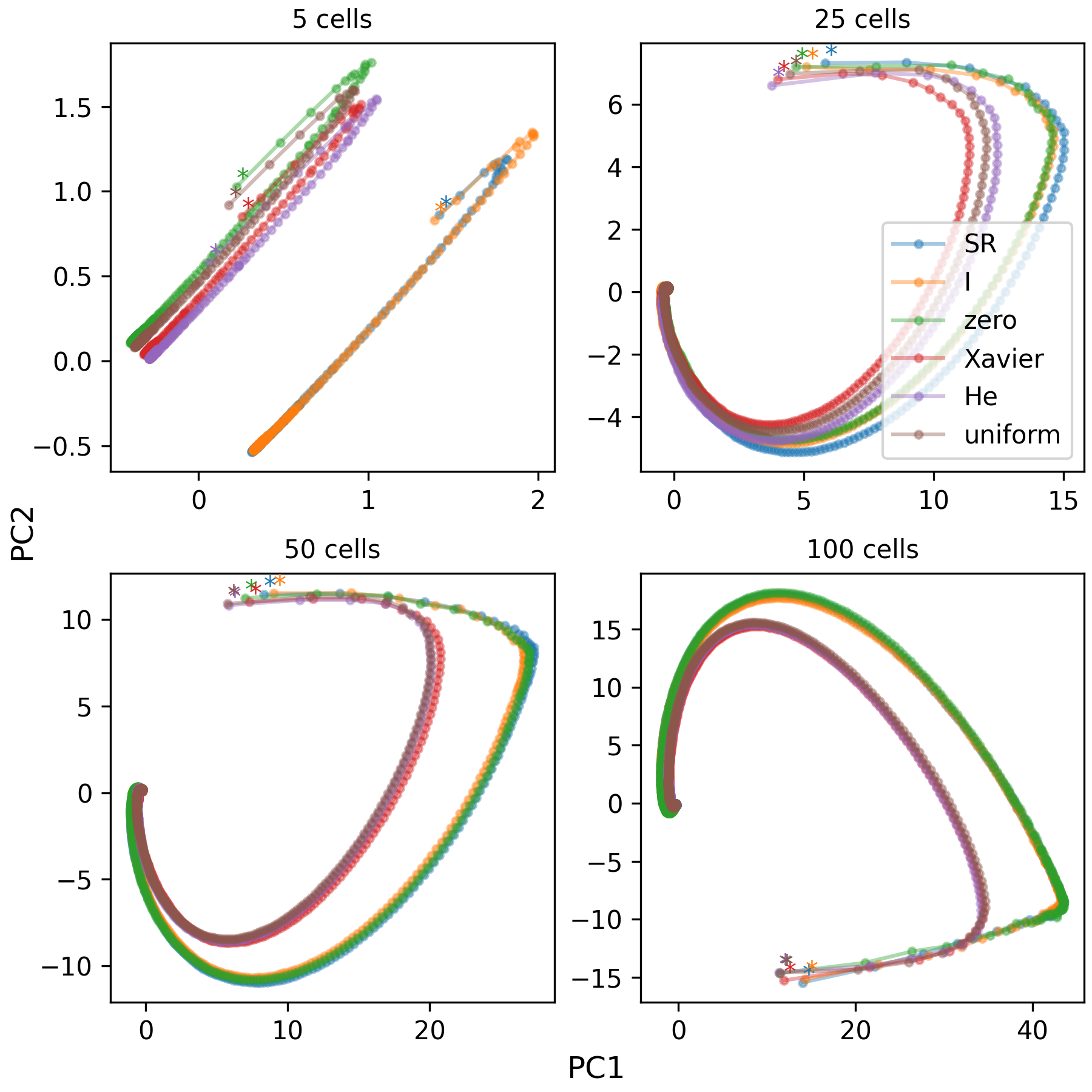}
		
		\caption{Principal component analysis (PCA) of SR place field matrix learning history shows that SF agents with random weights takes shorter route to converging optima. The simulated results from four different sizes of grid worlds (N=5, 25, 50, 100) are shown. Each dot shows PCA results of SR place fields after each episode (*, first episode).  Each line shows historical route of SR place field learning from SR or SF agents (SR, blue; SF weight initialization with identity matrix, orange; zero matrix, green; the Xavier method, red; the He method, purple; the uniform distribution, brown). The average of the SR place field matrices from 10 simulations was used for PCA.}
	    \label{PCA}
	\end{figure}

\subsubsection{Analyzing SR Matrix Changes with PCA}
The comprehensive SR matrix embodies the combined responses of all place cells, thereby collectively representing the entirety of the grid world. Consequently, to analyze and monitor the alterations in the learning pattern across episodes, a dimensionality reduction of the SR matrix is crucial. Borrowing methods from neuroscience research that are employed to examine large-scale neuronal recordings \cite{Cunningham:2014ev}, we utilized Principal Component Analysis (PCA) for this purpose. 

As hypothesized, and in alignment with the earlier observed transformations in the SR place field, our findings revealed that agents initialized with random weights followed a more direct path towards convergence (Figure \ref{PCA}).

In the case of a relatively compact grid world ($N=5$), a similar progression pattern in the SR place matrix was noticeable between the SR agents and SF agents initialized with identity matrices (the upper left panel of Figure \ref{PCA}). On the same note, agents initialized with random weights and SF agents with zero matrix initialization demonstrated analogous navigation patterns. These patterns, however, start to diverge with the expansion of the grid world ($N=25$). Here, random agents appear to emulate each other's trajectories, just as non-random agents do (the upper right panel of Figure \ref{PCA}). As we delve into larger grid worlds, for instance, $N=50$ and $N=100$, the distinctions between the random and non-random agents become increasingly apparent (the lower panels of Figure \ref{PCA}). It is in these settings that agents initialized randomly show a propensity for shorter convergence paths.

It is important to note that there are differences in the scale of the axes, and for the same axis scale, readers are referred to Fig. S\ref{supplePCA} in the supplementary materials.

	\begin{figure}[htp]
		\centering
		\includegraphics[width=3.5in]{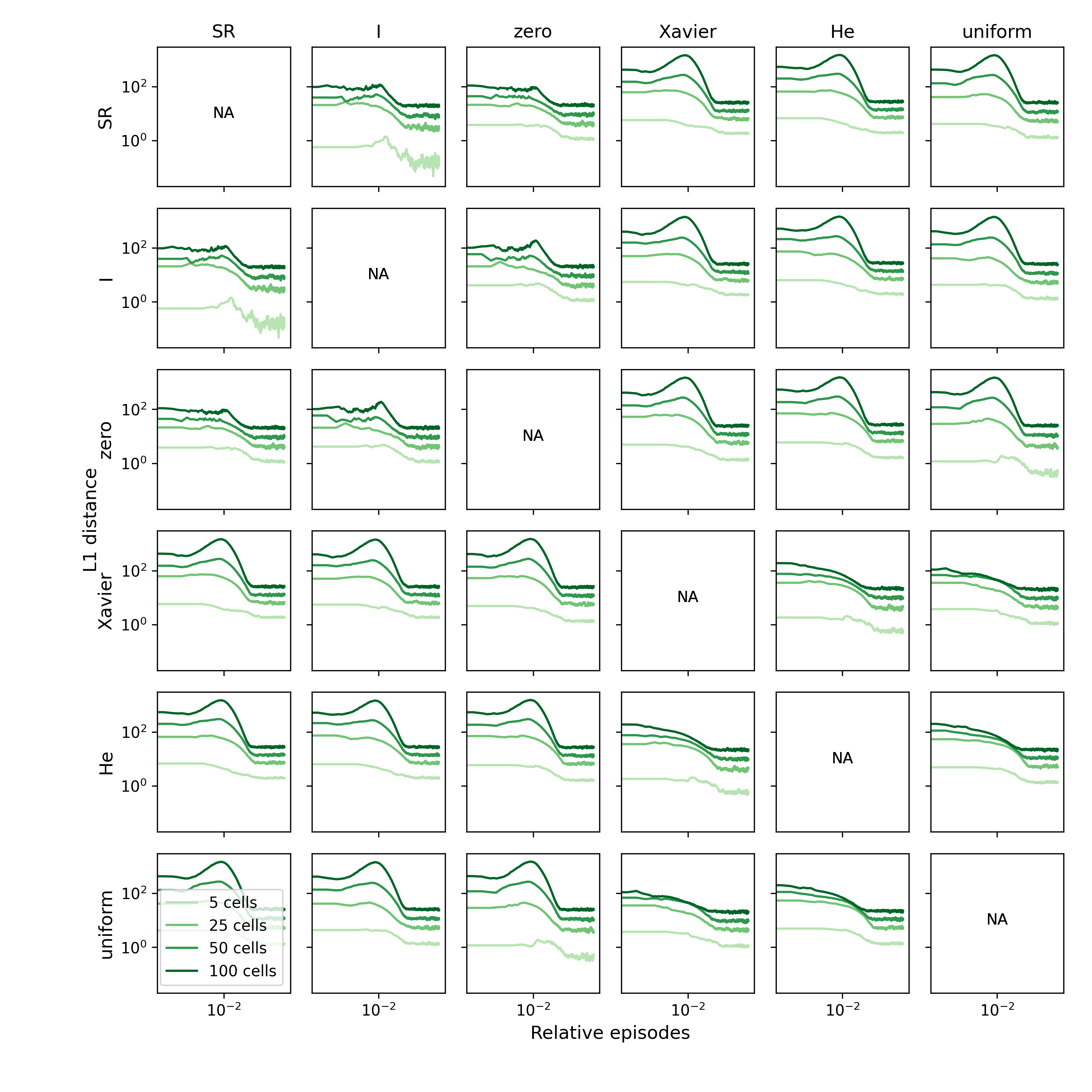}
		
		\caption{L1 distance between SR place fields of agents shows that SF agents with random weights differ from other non-random agents. Line plots show the change of L1 distance according to episode in four different sizes of grid worlds (N=5, 25, 50, 100). Since the total number of episodes depends on the grid world size, the relative episodes ($\frac{\text{episode}}{\text{total number of episodes}}$) are shown on the x-axis.}

	    \label{distance}
	\end{figure}

\subsubsection{Inter-agent SR Matrix Distance}

The PCA results suggest an intriguing possibility: if the randomly initialized agents achieve faster convergence towards the optimal SR place field, the distance between their SR place matrices and those of non-random agents should escalate during the learning process, eventually plateauing upon convergence. Conversely, the distance between the SR place matrices of non-random agents would remain relatively constant. 

To investigate this possibility, we computed the L1 distances between the SR matrices of the agents, presenting the results as a function of learning episodes (Figure \ref{distance}). The trends in the L1 distances over episodes among the six agents support our prediction, with the distance between random and non-random agents initially increasing before decreasing once again. Conversely, there's no discernible increase in the L1 distance when comparing either within the group of random agents or the group of non-random agents. Owing to the larger SR matrices and consequently larger distances found in larger grid worlds, all y-axes in Figure \ref{distance} are plotted on a logarithmic scale.

To mitigate the influence of the SR matrix's size on the L1 distance, we can divide the L1 distance by the total number of elements in the matrix ($N \times N$). This normalization procedure brings the metric down to the level of a single SR matrix element and further emphasizes that the distance between the randomly initialized agents and the non-random ones tends to increase (Fig. S\ref{suppleDist}).

\subsection{Enhanced Value Error and Step Length Reduction in Random Agents}

Drawing on Equations (\ref{value_decomposed}) and (\ref{successor_feature}), a direct correlation can be established between the variance in the SR matrix learning and the RL agent's performance. Herein, we analyze the anticipated value and step length to highlight the performance differences in the learning process.

	\begin{figure*}[!t]
		\centering
		\includegraphics[width=7in]{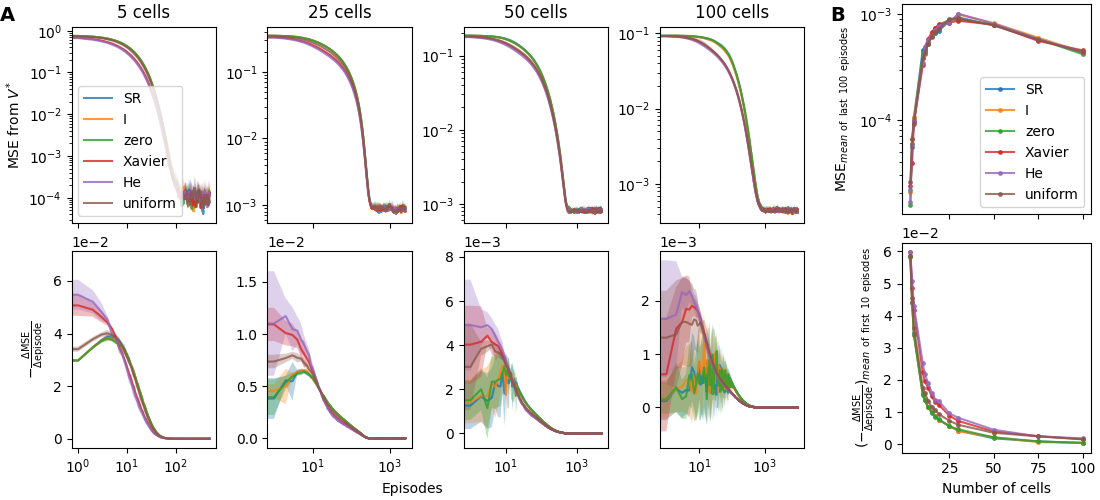}
		
		\caption{The mean square error of values shows that the estimated value of the SF agents with random weights decreases to the true value faster than the non-random agents. (A) The upper panel shows the mean squared error (MSE) of the estimated values ($V^{\pi}$) decreases as the episode progresses. The lower panel shows the decrease in the MSE per single episode ($-\frac{\Delta\textrm{MSE}}{\Delta\textrm{episode}}$). The results are from 10 simulations in four grid worlds of different sizes (arranged by columns). The averages (lines) and standard deviation (shades) of the SR or SF agents (SR, blue; SF weight initialization with identity matrix, orange; zero matrix, green; the Xavier method, red; the He method, purple; the uniform distribution, brown) are shown. (B) The upper panel shows that the averages of the MSE from last 100 episodes are similar across the SR or SF agents. The lower panel shows that the average of $-\frac{\Delta\textrm{MSE}}{\Delta\textrm{episode}}$ from first 10 episodes of the SF agents with random weights are larger than the non-random agents. Each circle marker indicates the size of grid worlds, which were simulated.}
	    \label{value_error}
	\end{figure*}
	
\subsubsection{Examination of Mean Square Error Decline Rate}

Taking into consideration the ground truth value ($V^{*}$), the mean squared error (MSE) of the estimated value ($V^{\pi}$) was calculated (please refer to Equation (\ref{V_error}) in section \ref{sec:eval}). Consistent with our expectations, we observed that the MSE of $V^{\pi}$ for the random agents diminished at a faster pace than for the non-random agents (the upper panel of Figure \ref{value_error}A).

We examined the rate of MSE decrement, represented as $-\frac{\Delta\textrm{MSE}}{\Delta\textrm{episode}}$ (the lower panel of Figure \ref{value_error}A). Early episodes, up to the 10th, illustrated a higher decrement rate in the randomized agents, indicating a more rapid reduction of the MSE value. It is noteworthy that among the random agents, those initialized utilizing the He and Xavier methods depicted a steeper reduction relative to those initialized uniformly (the lower panel of Figure \ref{value_error}A and B).

Nevertheless, in the latter episodes, the rate of MSE reduction exhibited minimal variation across the agents, underlining the comparable efficiency of the initialization methods in the long run.

	\begin{figure*}[!t]
		\centering
		\includegraphics[width=7in]{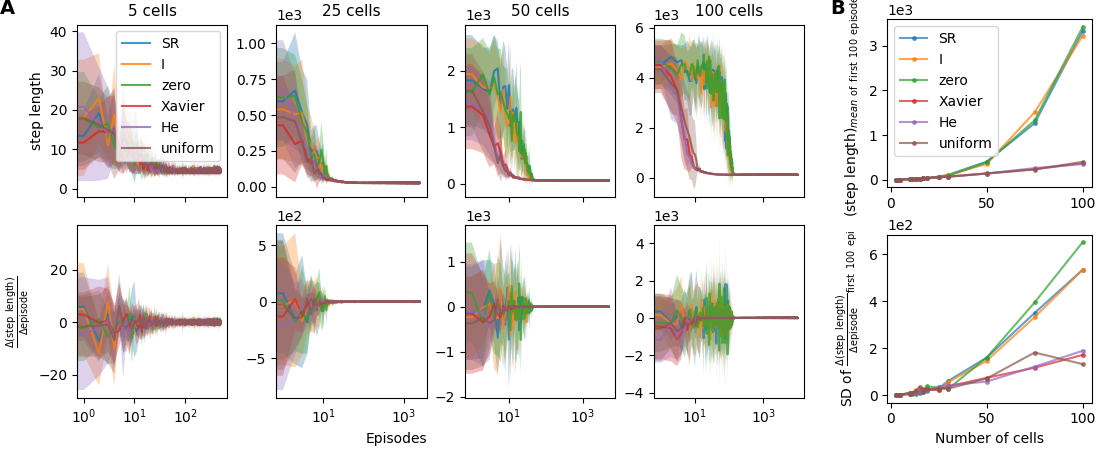}
		
		\caption{The SF agents with random weights converges to the optimal step length more rapid and stable than the non-random agents. (A) The upper panel displays the total step length taken to reach the target state for each episode. In the simulation results from the large grid world ($N=100$), the step length of the SF agents with random weights decreases to the ideal step length for the first 10 episodes, while the non-random agent decreases to the ideal step length after hundreds of episodes. The lower panel shows the decrease in total step length with each episode($\frac{\Delta(\textrm{step legnth})}{\Delta\textrm{episode}}$). In the simulation results of the large-scale grid world ($N \geq 50 $), the jittering of $\frac{\Delta(\textrm{step legnth})}{\Delta\textrm{episode}}$ of the SF agents with random weights disappears after 10 episodes, whereas its jittering of the non-random agents persists. The results are from 10 simulations in four grid worlds of different sizes (arranged by columns). The averages (lines) and standard deviation (shades) of the SR or SF agents (SR, blue; SF weight initialization with identity matrix, orange; zero matrix, green; the Xavier method, red; the He method, purple; the uniform distribution, brown) are shown. (B) The top panel shows the average step length of the first 100 episodes according to grid world size. The lower panel shows the standard deviation of $\frac{\Delta(\textrm{step legnth})}{\Delta\textrm{episode}}$ of the first 100 episodes according to grid world size. }
	
	    \label{step_length}
	\end{figure*}

\subsubsection{Step Length Reduction}

Given the quick reduction observed in the MSE of random agents, we can anticipate a corresponding accelerated decline in the step length to the goal cell within each episode of the grid world exploration. Due to their initially high $\epsilon$ probability, all RL agents undertake an exploration of the grid world that mimics a random walk, which naturally results in longer step lengths during the early episodes. As shown in Figure \ref{step_length}, as the exploration episodes advance, the step length predictably shrinks to the size of the grid world.

In smaller grid worlds (where $N<30$), no significant differences in the reduction of step lengths amongst the RL agents were observed. However, as the grid world's size expands ($N>=50$), it was noted that the step length of random agents diminished at a faster rate (see Figure \ref{step_length}B, upper). 

The trajectory of step length reduction manifested clear distinctions between the two groups. Non-random agents demonstrated significant fluctuations in the decrement of step length, while such variance was less prevalent in random agents. This disparity was further illustrated by calculating the rate of step length reduction, $\frac{\Delta(\textrm{step length})}{\Delta\textrm{episode}}$, and evaluating its standard deviation (the lower panel of Figure 6A and B).

For non-random agents, the fluctuations in the rate of step length reduction inflated exponentially with the increase in grid world size. Conversely, for the random agents, the fluctuations displayed a linear growth pattern despite the expanding grid world size, indicating a more stable decrease in step length as the learning process progressed.

\section{Discussion and Conclusion}

In this study, we investigated the role of initial weight matrix configurations in the efficiency of SF learning. We scrutinized three initialization methods: the identity matrix, zero matrix, and random matrix (using Xavier, He, and uniform distribution). Our results demonstrated that the randomized agents, regardless of the specific initialization method, outperformed the identity and zero matrix agents. Specifically, we found that the random agents learned faster, which was evident from the decrease in MSE of the estimated value and step length to the goal cell in a grid world environment. Further PCA analysis illuminated the distinct patterns of learning in randomized agents compared to non-randomized ones, which provided additional insight into the evolution of SR place matrix. Thus, our findings underscore the significant influence of initial weight configurations on the effectiveness and speed of SF learning.

\subsection{Interpretation of SF Weight Matrix Initialization}

Initiating the SF weight matrix as an identity matrix provides the agent with a unique starting position in its learning journey about the environment. As learning progresses, each element in the identity matrix corresponds to a particular state, thereby facilitating the updating of knowledge regarding state transitions. However, this initialization method could restrict the agent's versatility in exploring and learning diverse environmental patterns, potentially resulting in slower learning as observed in previous research. This limitation could be particularly consequential for an agent's adaptability in increasingly complex or dynamic environments.

Alternatively, initializing the SF weight matrix with zero establishes a "tabula rasa" situation for the agent. Devoid of any prior knowledge, these agents are heavily influenced by their environmental interactions and the inherent learning algorithm. Although this approach broadens the exploration scope, it may decelerate learning due to the absence of initial guidance. This downside was apparent in studies where agents initiated with a zero matrix took a longer time to converge compared to their randomly initialized counterparts.

Contrastingly, random matrix initialization strikes a balance between exploration and exploitation. Incorporating random elements into the SF weight matrix equips the agent with a degree of "innate knowledge" guiding its initial steps while preserving a vast spectrum for exploration and learning. Consequently, this initialization method may enhance the learning efficiency, offering a promising avenue for improving SF learning algorithms.

\subsection{Impact of Xavier and He Initialization Methods on Agent Learning in MDP-Based Models}

Effective initialization methods, contingent upon the activation function, have been well-documented in the study of Artificial Neural Networks (ANN) utilizing backpropagation algorithms. The normalized Xavier initialization method \cite{glorot2010} is typically employed with sigmoid and tanh functions, while the He initialization method \cite{he2015} sees frequent usage with ReLU functions.

In this study, the absolute values derived from the Xavier and He methods were employed to initialize the random agent. It's worth noting that inclusion of negative numbers in the weight matrix can result in a negative SR value corresponding to future occupancy, as we make use of a single-layer function approximator devoid of an activation function. To address this issue, we can utilize multilayer ANNs as a function approximator. When a deep neural network is employed as an SF approximator, it begs the question as to which activation function in the hidden layer is optimal, and consequently, the most effective weight initialization method.

Though this paper focused on exploring MDP-based agent learning of environmental characteristics via the SF algorithm, numerous other algorithms are available that describe animal-environment interactions and learning mechanisms. For instance, Particle Swarm Optimization (PSO) that models avian foraging behavior \cite{eberhart,kennedy}. Among the latest advancements to the PSO algorithm, multi-swarm PSO has been successfully implemented in feature learning for sentiment analysis of Massive Online Open Course lecture reviews \cite{liu2016}.

\subsection{Neurobiological Considerations}

In the context of RL, the feature vector of the input layer offers a snapshot of the agent's current position. Subsequently, this information is transformed by the SF weight matrix into a population vector, effectively encoding the anticipated future occupancy given the policy at hand. This sequence of operations bears striking resemblance to the neurobiological mechanisms believed to underpin spatial learning.

A collection of studies \cite{decothi2020, geerts2020, stachenfeld2017} suggest that hippocampal CA1 place cells encode SR through population codes. Viewed through this neurobiological lens, the SF weight matrix may be interpreted as a close analog to the synaptic weights connecting CA1 place cells to preceding layers of neurons in the neural hierarchy, such as those in the CA3 and entorhinal cortex. This parallel between the functioning of RL algorithms and the neuronal processes that facilitate spatial learning lends support to the use of such algorithms in the investigation of cognition and its underlying biological substrates.

While the exact mechanisms by which the brain might implement the synaptic update rule used in our study remain elusive, a body of research has found substantial evidence that TD learning parallels the activity of dopaminergic neurons in response to reward prediction errors \cite{montague1996, schultz1998}. This aligns with the hypothesis that the neural instantiation of TD learning might be facilitated through neuroplasticity rules, such as spike timing-dependent plasticity and heterosynaptic plasticity \cite{lee2020a, rao2001:nc}. This conjecture, if further corroborated, could add an extra layer of understanding to our exploration of the intersections between artificial intelligence and neurobiology.

From a biological standpoint, it seems reasonable to posit that place coding and reward prediction coding might be processed in tandem within the brain, which subsequently synthesizes these elements into anticipated values for a given state. This line of thought supports the perception of the brain as a device engaging in parallel distributed processing, as suggested by \cite{rumelhart1986}. Underpinning this proposition, the backpropagation algorithm has exhibited exceptional capability in tasks such as image recognition \cite{krizhevsky2012, Rumelhart:1986er}. Moreover, a convolutional neural network (CNN) trained with this algorithm has exhibited activation patterns that bear resemblance to those observed in the visual cortex and the inferior temporal cortex of the brain \cite{Yamins:2014gi}. Notably, when the activation pattern of a trained CNN was used to manipulate an image, it was found to predict neuronal responses in the V4 visual cortex of macaque monkeys \cite{bashivan2019}. Nevertheless, it is still a matter of ongoing debate and remains unconfirmed whether the backpropagation algorithm is genuinely operative within the brain \cite{lillicrap2020, whittington2019}.

While the biological embodiment of SR learning remains elusive, particularly regarding the brain's processing location and method for the inner product of the feature vector and reward vector, there is a notable correlation between the outcomes of SR learning and the behavior of hippocampal place cells \cite{gershman2018, momennejad2017, stachenfeld2017}. However, it warrants further exploration to fully understand how the brain learns and signifies the sequences of state transitions, rewards, and state values. Experimental findings have associated the representation of the reward signal with the orbitofrontal cortex (OFC) \cite{gottfried2003, sul2010}, suggesting the anterior cingulate gyrus as a probable area for the integration of the OFC's reward signal and the HPC's SF signal \cite{shenhav2013, kolling2016}. In contrast, a study by \cite{gauthier2018} postulated that the HPC directly encodes the position of the reward.

Transitioning our focus to the question of 'how', we are confronted with the challenge of extracting a scalar value from the successor feature vector and reward vector \cite{meyniel2015}. Although this issue extends beyond the scope of the current study, we can glean some neurobiological insights. Specifically, if the synaptic weights in a neural network at the developmental stage are randomly initialized, they demonstrate faster convergence to an optimal state.

\subsection{Limitation of the study}

While our study offers valuable insights into the impact of SF weight matrix initialization on learning efficiency and convergence, it's important to note that these findings are based on a one-dimensional grid world. The learning patterns and efficiency we observed may vary with different types of environments. For instance, we noticed more distinct learning trajectories in larger grid worlds, while smaller grid worlds exhibited similar patterns. Therefore, the grid world size is a potential limitation of our study, and our conclusions might be more applicable to larger grid worlds. Future studies could broaden the scope by investigating a wider range of MDP environments, such as two-dimensional grid worlds, to enhance the applicability of our findings.

Our study relied on certain evaluation metrics, including MSE of value error, step length, and PCA of SR place matrix, to analyze learning efficiency and convergence. While these metrics provided significant insights, they might not encapsulate all aspects of an agent's learning trajectory. For instance, MSE and step length predominantly focus on the speed of learning, potentially overlooking other critical dimensions such as stability and adaptability of learning. Additionally, PCA, while effective in dimensionality reduction and visualizing high-dimensional data, may oversimplify complex learning patterns.

\subsection{Conclusion}
This study embarked on an exploratory journey into the role of matrix initialization in SF learning within the framework of RL. We discovered notable differences in the learning trajectories of agents with different matrix initialization forms - identity, zero, and random (Xavier, He, uniform distribution). Our findings suggest that random matrix initialization, particularly using Xavier and He methods, led to more efficient learning and faster convergence to the optimal state, as evidenced by a quicker decrease in value error and step length. The PCA further revealed distinct patterns of SR place matrix evolution among different agents, reinforcing the importance of matrix initialization in shaping learning dynamics.

The study highlights the significance of weight initialization in the learning process. Our observations demonstrate that the choice of initialization method significantly influences the learning trajectory and efficiency of the agents. Specifically, agents initialized with random matrices demonstrated accelerated learning and quicker convergence to the optimal state. These findings underline the value of exploring diverse initialization techniques to enhance the effectiveness of SF learning.

The implications of this research extend beyond SF learning and RL, contributing to our broader understanding of intelligence from both a neuroscientific and artificial intelligence perspective.  By drawing parallels between SF learning and the functioning of place cells in the brain, the study offers intriguing insights into the neurobiological processes underlying learning. An intriguing direction for future research is to delve deeper into the parallels and disparities between biological learning and AI learning algorithms. The results of this study shed light on the learning efficiency of agents, mirroring the learning process of place cells in the brain. Continued efforts to bridge this gap could lead to the development of more biologically-inspired AI models, possibly leading to breakthroughs in our understanding of both artificial and natural intelligence.

\section*{Acknowledgments}
This study was supported by the National Research Foundation of Korea(NRF) grant funded by the Korea government(MSIT; Ministry of Science and ICT)(No. NRF-2017R1C1B507279). The author would like to thank ChatGPT for their assistance in editing and improving the language of the paper, as well as for their helpful brainstorming sessions.

\bibliographystyle{IEEEtran}
\bibliography{SF_W_ieee.bib}

\renewcommand{\section}[2]{}

\renewcommand{\figurename}{Supplementary Figure}
\setcounter{figure}{0}

\begin{figure*}[!htb]
	\centering
	\includegraphics[width=5in]{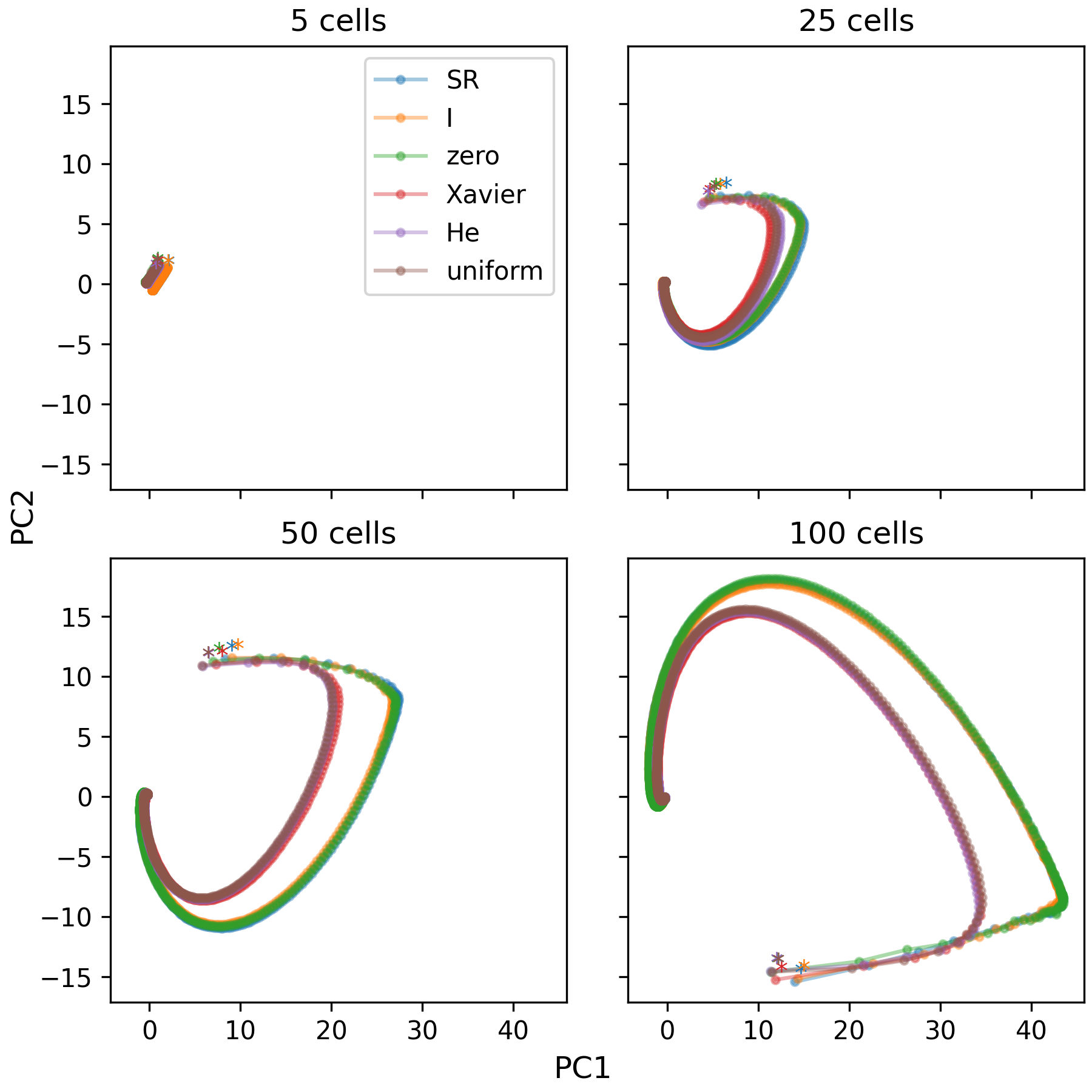}
	
	\caption{The same principal component analysis (PCA) of the SR place field matrix learning history as shown in Figure 3, but drawn to the same scale. Except for the scale, the details are the same as in Figure 3.}
    \label{supplePCA}
\end{figure*}

\newpage 

\begin{figure*}[!htb]
	\centering
	\includegraphics[width=5in]{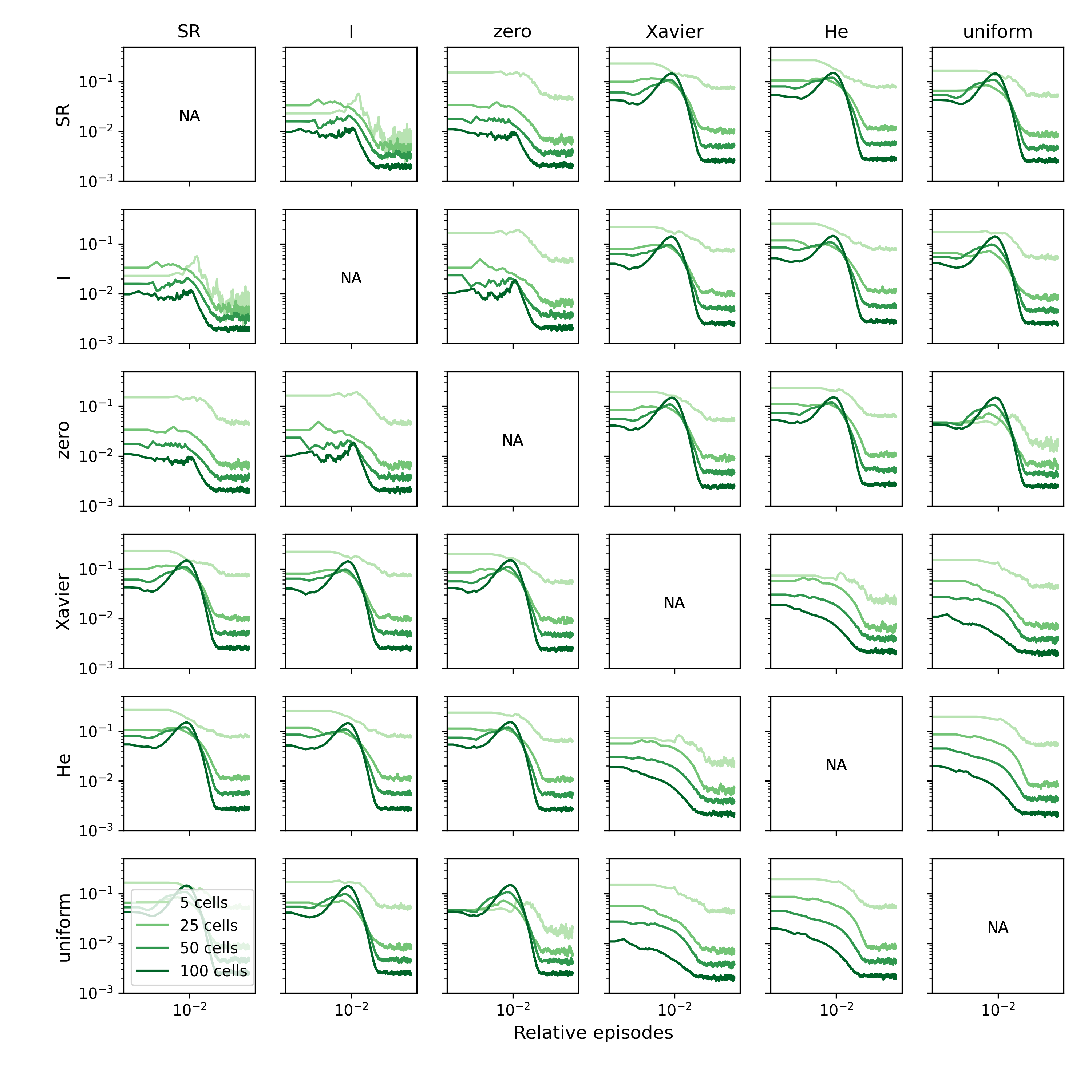}
	
	\caption{L1 distances divided by the size of the matrix ($N \times N$) are shown. This normalization shows the distance between one element of the SR matrix. Random agents has shown great distance from the non-random agents.}
    \label{suppleDist}
\end{figure*}

\newpage

\begin{IEEEbiography}[{\includegraphics[width=1.5in,clip,keepaspectratio]{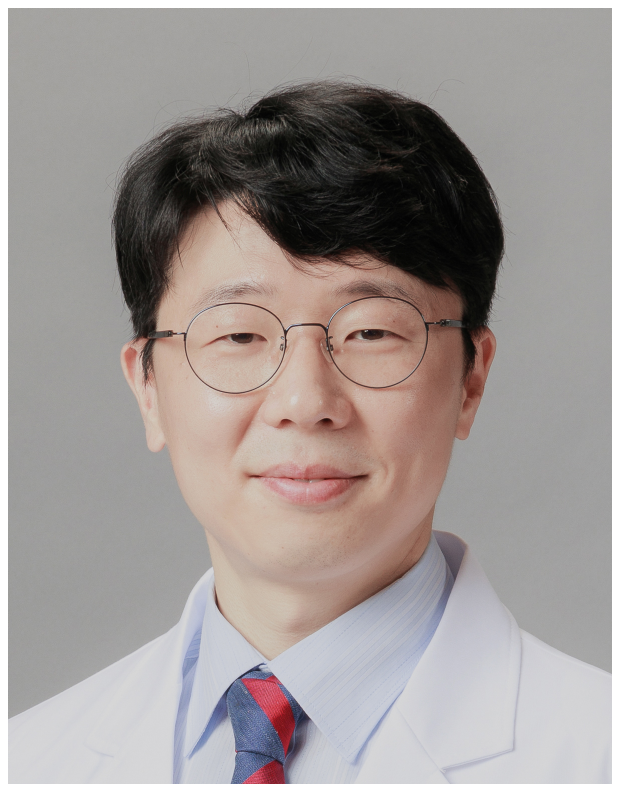}}]{Hyunsu Lee,} a dynamic and innovative Assistant Professor, is currently utilizing his expertise in neuroscience at Pusan National Unversity, School of Medicine to further his cutting-edge research in the field of artificial intelligence. Driven by a passion for exploring the intersection of neuroscience and machine learning, his current research endeavors are focused on the medical and neuroscientific applications of reinforcement learning algorithms.
\end{IEEEbiography}

\vfill

\end{document}